# Insights into muscle metabolic energetics: Modelling muscle-tendon mechanics and metabolic rates during walking across speeds

Israel Luis[1]*, Maarten Afschrift[2], Friedl De Groote[3], Elena M. Gutierrez-Farewik[1,4]

[1] KTH MoveAbility Lab, Dept. Engineering Mechanics, KTH Royal Institute of Technology, Stockholm, Sweden

[2] Faculty of Behavioural and Movement Sciences, VU Amsterdam, Amsterdam, The Netherlands

[3] Department of Movement Sciences, KU Leuven, Leuven, Belgium

[4] Department of Women's and Children's Health, Karolinska Institutet, Stockholm, Sweden

* Corresponding author, email: ailp@kth.se

## Abstract

The metabolic energy rate of individual muscles is impossible to measure without invasive procedures. Prior studies have produced models to predict metabolic rates based on experimental observations of isolated muscle contraction from various species. Such models can provide reliable predictions of metabolic rates in humans if muscle properties and control are accurately modeled. This study aimed to examine how muscle-tendon model calibration and metabolic energy models influenced estimation of muscle-tendon states and time-series metabolic rates, to evaluate the agreement with empirical data, and to provide predictions of the metabolic rate of muscle groups and gait phases across walking speeds. Three-dimensional musculoskeletal simulations with prescribed kinematics and dynamics were performed. An optimal control formulation was used to compute muscle-tendon states with four levels of individualization, ranging from a scaled generic model and muscle controls based on minimal activations, to calibration of passive muscle forces, personalization of Achilles and quadriceps tendon stiffnesses, to finally informing muscle controls with electromyography. We computed metabolic rates based on existing models. Simulations with calibrated passive forces and personalized tendon stiffness most accurately estimate muscle excitations and fiber lengths. Interestingly, the inclusion of electromyography did not improve our estimates. The whole-body average metabolic cost was better estimated using Bhargava et al. 2004 and Umberger 2010 models. We estimated metabolic rate peaks near early stance, pre-swing, and initial swing at all walking speeds. Plantarflexors accounted for the highest cost among muscle groups at the preferred speed and was similar to the cost of hip adductors and abductors combined. Also, the swing phase accounted for slightly more than one-quarter of the total cost in a gait cycle, and its relative cost decreased with walking speed. Our prediction might inform the design of assistive devices and rehabilitation treatment. The code and experimental data are available online.

## Author Summary

Simulations of the musculoskeletal system can provide insights into motion mechanics and energetics. To increase confidence in the approach, it is important to establish how predictions are affected by modeling assumptions. In this study, we evaluated how simulations with increasing levels of sophistication and metabolic energy models from the literature agreed with recorded muscle excitations, fiber lengths, and whole-body metabolic rates during walking at different speeds. We found that calibrating muscle passive components: muscle passive forces, and tendon stiffness, estimated better muscle excitations and fiber lengths than using original musculoskeletal model values. The metabolic energy cost of walking was closely estimated with only a subset of metabolic energy models. We estimated that the metabolic energy demands are the highest during three periods of the gait cycle - early stance, pre-swing, and initial swing, at all walking speeds. In all walking speeds, muscles that support plantarflexion accounted for the highest metabolic cost among all the muscle groups. As walking speed increases, the metabolic cost of the stance phase also increases. Our detailed description of metabolic costs at muscle groups and gait phases might help to design assistive devices and rehabilitation treatments.

## Keywords

# 1. Introduction

Metabolic rate is a measure of the metabolic demand of muscular effort during motion and basal functions over time. A comprehensive understanding of the metabolic rate can support rehabilitation treatments (1) or inform the design of assistive devices (2,3). Whole-body average metabolic rate is typically measured indirectly through spiroergometry, whereby oxygen and carbon dioxide rates in inhaled and exhaled breaths are measured as a function of time (4). This measurement only provides a metabolic assessment in constant-intensity conditions, as slow mitochondrial dynamics, body transit delays, respiratory control mechanisms, and high breath-by-breath variability make instantaneous metabolic rate mapping impossible (5). For instance, measuring average metabolic rates of walking requires spiroergometric measurements for 5-6 minutes at a given speed (5). State-of-the-art wearable systems can accurately predict instantaneous metabolic rates, but only at the whole-body level (6). Methods to measure the contribution of individual muscle metabolic rates are available, though limited to experimental animal studies (7). An alternative approach to estimating individual muscle metabolic rates is using computational methods such as musculoskeletal modeling.

Musculoskeletal simulations can predict individual muscle metabolic rates using muscle-tendon and metabolic energy models. The muscle-tendon unit of the musculoskeletal system is typically described based on the Hill-type mechanical model. This model consists of a contractile and a parallel elastic element representing the muscle's active and passive force-generation capacities, respectively, and a serial elastic element representing the tendon and aponeurosis (8). The Hill-type model does not describe the biochemical interactions involved in cross-bridge cycling nor metabolic cost but rather the dependency between force production and muscle activation, fiber length, and fiber contraction speed. In this regard, metabolic energy models have been developed to predict the metabolic rate of muscles from the contractile element work rates and heat rates derived from the muscle-tendon states (9–11).

Multiple metabolic energy models have been developed based on experimental datasets and assumptions of muscle energetics. Several models have been used frequently, including those proposed by Bhargava et al. (BH04) (9), Houdijk et al. (HO06) (12), Lichtwark and Wilson in its original (LW05) and modified version (LW07) (13), Umberger in its initial (UM03) (11) and revised version (UM10) (14), and a further modification proposed by Uchida et al. (UC16) (15) (Table 1). These models use the same definition of contractile element work rate but different formulations of the heat rate. The total heat rate is primarily composed of four terms: activation rate, associated with non-contractile costs of muscle activation to pump Ca++ ions against a concentration gradient across the sarcoplasmic reticulum; maintenance rate, associated with the cycling of actin-myosin cross-bridges during isometric tetanus; and shortening and lengthening rates, associated with an extra cost of actin-myosin cross-bridge interactions during dynamic muscle contractions.

Metabolic energy models have been reported to predict metabolic rates during walking in people with and without motion disorders and assistive devices with reasonable accuracy. The whole-body average metabolic rate of walking from forward dynamics simulations was estimated as 5.8 W/kg at 1.36 m/s with the BH04 model (9), 4.4 W/kg at 1.2 m/s with the UM03 model (11), ~5 W/kg at 1.3 m/s with the UM10 model (14), and 4 W/kg at 1.46 m/s using the UC16 model (16). These estimations were relatively close to experimental energy measured with indirect calorimetry, except with the BH04 model, which was higher. Interestingly the BH04 model accurately estimated metabolic rates in a small clinical population, specifically in two subjects post-stroke across slow walking speeds (17) and in an amputee with a transtibial passive prosthesis at 1.3 m/s (18). The metabolic rate changes related to using assistive devices in able-body individuals during hopping were estimated with the LW07 model (19) and during walking with the UC16 model (20). Despite these encouraging findings, evaluating the accuracy of various metabolic energy model formulations is challenging, as they depend on the underlying muscle-tendon states, which are in turn dependent on musculoskeletal models and simulation approaches.

Two previous studies have assessed the accuracy of and differences between metabolic energy models in both forward simulations and inverse dynamic analyses. Miller compared metabolic energy models in tracking and predictive simulations at 1.45 m/s and reported that none predicted the absolute value of (3.35 $Jm^{-1}kg^{-1}$) accurately (21). Their estimates varied from 2.45 $Jm^{-1}kg^{-1}$ using the BH04 model to 7.15 $Jm^{-1}kg^{-1}$ using the LW07 model. Miller also found that differences in the metabolic energy models are mainly attributed to descriptions of active lengthening and eccentric work. Koelewijn et al. performed 2D musculoskeletal simulations with prescribed kinematics and dynamics and evaluated walking at two speeds and three incline levels. They reported a good correlation between estimated metabolic cost and indirect calorimetry measurements, though the estimated cost tended to underestimate the recorded cost, particularly at higher metabolic demands (22).

Further insights may be gained by evaluating the metabolic energy models based on as accurate estimations of the muscle-tendon states as possible. Simulations performed by Miller (21) and Koelewijn et al. (22) estimated muscle-tendon conditions by assuming that human motor control in movement minimizes muscle effort. They also validated muscle control by comparing muscle activations to electromyography signals (EMGs). Nonetheless, further validation of the muscle-tendon states might be required to gain confidence in estimating metabolic rates. We have recently evaluated agreements between estimated muscle activations and fiber lengths with experimental observations (23). We found that whereas variations of musculoskeletal models estimated similar muscle activations that agreed with recorded EMGs, the operating ranges of modelled muscle-tendon actuator contractions varied among models. The operating range affects the force generation of the muscle and thus affects mechanical work and metabolic rate (24). Accurate models of muscle-tendon mechanics, including tendon compliance (13), passive forces (25), and muscle control (26), are suggested to play a critical role in the estimation of muscle force during walking and, therefore, are likely to influence the metabolic rate of muscles. Previous comparative studies did not validate their estimates of muscle fibers nor added information from EMGs into the muscle control, which might provide more reliability in predicting metabolic rates.

In this study, we used musculoskeletal simulations informed by various degrees of experimental observations and metabolic energy models to illustrate how multiple degrees of individualization influenced our predictions of muscle-tendon states and, thus, muscle metabolic rates. We evaluated the accuracy by comparing the estimates of muscle activations, fiber lengths, and metabolic cost with experimental observations recorded from our study or obtained from the literature. Then, based on the simulation with the highest accuracy, we aimed to provide a reliable prediction of the metabolic cost of the gait phases and muscle function groups across walking speeds.

## 2. Methods

### 2.1. Participants

Eight unimpaired adults (5/3 male/female, age: 33.6 ± 8.5 years old, height: 1.74 ± 0.10 m, body mass: 71.3 ± 10.9 kg [mean ± SD]) participated in this study. The study was approved by the Swedish Ethical Review Authority (Dnr. 2020-02311), and all participants provided written consent. Participation was voluntary and could be terminated at any time during the experiment.

## 2.2. Experimental protocol and data processing

Subjects walked on a treadmill at 55%, 70%, 85%, 100%, 115%, 130%, and 145% of their estimated preferred walking speed (PWS) in two conditions: treadmill, then overground walking. The PWS was predetermined by the participant's gender, age, and height (27). In treadmill trials, oxygen, and carbon dioxide respiration rates were recorded during 6 minutes of walking at each speed (Cortex Metamax 3B, Leipzig, Germany). The representative value from each speed was computed based on the average in the last 3 minutes. After each trial, each subject rested for 5 minutes to avoid fatigue. At each experimental session, speed was randomized, and average cadence was recorded.

In overground walking experiments, participants walked along a lab pathway and emulated different walking speeds by matching recorded cadences from treadmill walking with audio signals from a metronome app. Marker position (100 Hz) and ground reaction forces (1000 Hz) were measured using optical motion capture (Vicon V16, Oxford, UK) and strain gauge force platforms (AMTI, Watertown, MA, USA), respectively. Full-body marker placement was implemented based on the Conventional Gait Model with the Extended-foot model (CGM 2.4). EMG from eight muscles in one leg were recorded: biceps femoris long head, semitendinosus, vastus lateralis and medialis, tibialis anterior, gastrocnemius lateralis and medialis, and soleus, using bipolar surface wireless electrodes (Myon aktos/Cometa systems, Milan, Italy). The selection of sides for EMG placement was randomized. Skin preparation and electrode placement followed the Electromyography for the Non-Invasive Assessment of Muscles guidelines (SENIAM) (28). Subjects were asked to perform functional tests to corroborate EMG placement. EMG signals were processed using a 4th-order zero-lag Butterworth band-pass filter (20-400 Hz), full-wave rectification, and a 4th-order zero-lag Butterworth low-pass filter (6 Hz). EMG of the vastus intermedius was estimated as an average of the vastus lateralis and medialis. Ground reaction forces were processed using a 4th-order zero-lag Butterworth low-pass filter (35 Hz).

## 2.3. Musculoskeletal simulation framework

### 2.3.1. Musculoskeletal model, scaling method, and inverse kinematics and dynamics

The generic musculoskeletal model developed by Rajagopal et al. (29) was selected for this study. We scaled the generic model using OpenSim's Scale Tool, which adjusted muscle paths (origins/insertions), skeletal geometry, and segment inertial properties to fit anthropometric dimensions obtained from a captured static calibration trial and scaled the optimal fiber lengths and tendon slack lengths of each muscle-tendon actuator linearly to preserve the ratio of the generic model in the scaled model. The maximum isometric force of each muscle-tendon actuator was individualized based on the expected muscle volume, which in turn was estimated based on body weight and height, according to a regression equation proposed by Handsfield et al. (30). Each muscle's maximum isometric force is assumed to be directly proportional to its physiological cross-sectional area and specific tension. Muscle active and passive force-length and force-velocity relationships were modeled as described by De Groote et al. (31).

Marker trajectories and ground reaction forces throughout 21 gait cycles per subject (3 cycles at each of 7 walking speeds) were analyzed with inverse kinematics and inverse dynamics using OpenSim 4.1. Marker tracking weights for inverse kinematics were selected to minimize the error between experimental and virtual markers. The subtalar and metatarsal joints were fixed at neutral anatomical positions.

These model and muscle-tendon scaling methods were selected based on a recent study in which we determined that this combination yielded the best agreement between computed fiber lengths and those measured with ultrasound and between computed muscle excitations and measured EMGs in the vastus lateralis, and medialis, tibialis anterior, gastrocnemius medialis and lateralis and soleus (23). In that study, we observed a poor estimation of the activations in the biceps femoris long head, semitendinosus, which we hypothesized to be due to the overestimation of passive tendon forces at knee and hip joints (29,32). We address this concern in the current study.

*2.3.2. Calibration of passive force-length relationships*

Parameters of muscle-tendon actuator passive force-length relationships were calibrated to better agree with experimental findings reported by Silder et al. (33). The relationship was modeled based on OpenSim's Thelen2003Muscle (34) (Equation 1)

$$f_{pas}(\tilde{l}_M) = \frac{e^{\frac{k_{PE}(\tilde{l}_M - s_0)}{s_M}} - 1}{e^{k_{PE}} - 1} \quad \text{(Eq. 1)}$$

Where $\tilde{l}_M$ is the normalized fiber length, $k_{PE}$ is the exponential shape factor for the passive force-length relationship, $s_0$ is the normalized fiber length at which the passive force starts to increase, and $s_M$ is the normalized fiber length, measured from the optimal fiber length ($\tilde{l}_M = 1$), at which maximum force is reached. We formulated an optimization routine wherein we selected $k_{PE}$ and $s_0$ as optimization variables, with the objective function of minimizing the error between the simulated joint moments from the model and the experimental joint moments reported by Silder et al. (33). The calibrated passive forces closely resembled the measured passive joint moments reported by Silder et al. (S1 Fig). Detailed information about the optimization routine is presented in Supplementary material A: Calibration of passive forces.

*2.3.3 Simulation workflow and optimized muscle-tendon parameters*

We computed muscle excitations, states, and state derivatives using four degrees of individualization in our simulations by (1) starting from the scaled generic model and muscle recruitment based on minimum effort, then consequently including (2) calibration of passive forces, (3) personalization of Achilles and quadriceps tendon stiffnesses, and (4) muscle controls informed with EMGs (Fig 1). The following simulations were performed in order of increasing complexity:

EFFORT-GEN: Minimal muscle effort with generic passive force. Muscle excitations were computed assuming optimal motor control that minimizes muscle activations squared.

EFFORT-CAL: Minimal muscle effort with calibrated passive force. Muscle excitations were computed assuming optimal motor control that minimizes muscle activations squared.

EFFORT-TEN: Minimal muscle effort with calibrated passive force and personalized tendon stiffness. Muscle excitations were computed assuming optimal motor control that minimizes muscle activations squared. Personalized Achilles and patellar tendon stiffness are as in EMG-TEN.

EMG-TEN: Muscle controls derived from minimal muscle effort and EMGs, with calibrated passive force and personalized tendon stiffness. Muscle excitations of the biceps femoris long head, semitendinosus, vastus lateralis and medialis, tibialis anterior, gastrocnemius lateralis and medialis, and soleus tracked the recorded time-series EMG data. Remaining muscle excitations were computed assuming optimal motor control that minimizes muscle activations squared. Thus, the objective function minimized muscle activations and the deviation between estimated excitations and EMGs. Achilles and quadriceps tendon stiffnesses were added as design variables and optimized using all the gait cycles simultaneously per subject.

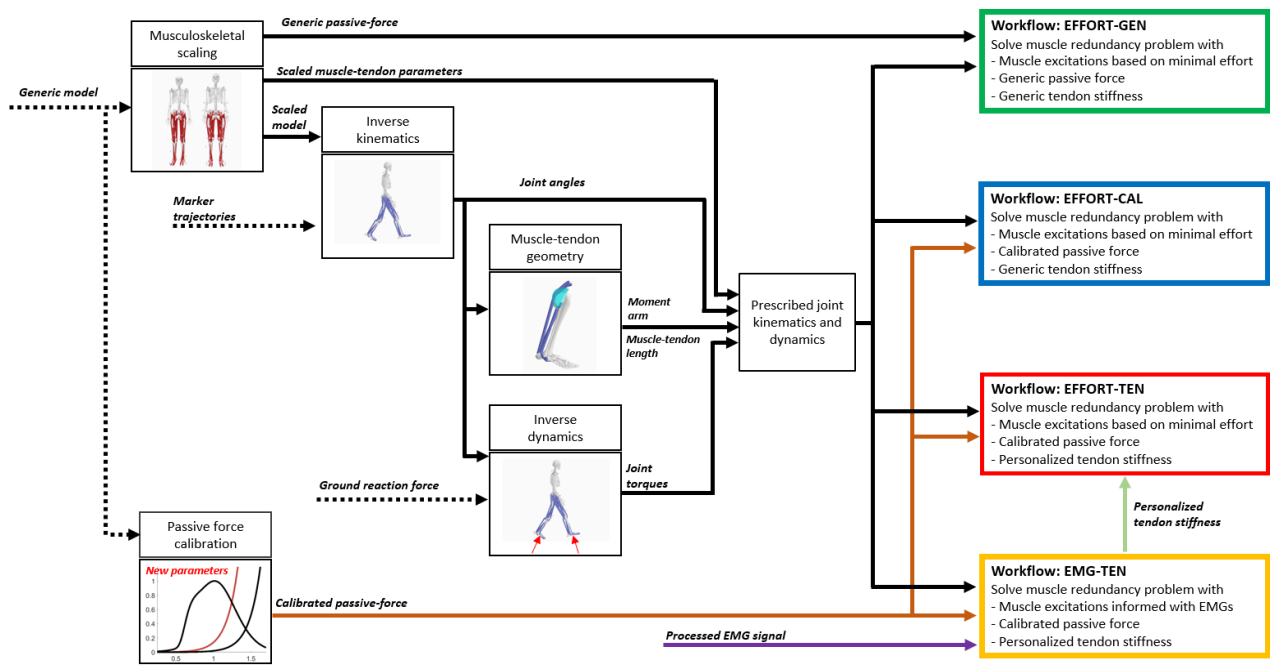

Fig 1: *Diagram of the simulation workflows*.

In all simulations, inverse kinematics and inverse dynamics solutions were prescribed. Also, muscle excitations were estimated by minimizing activations squared while imposing that the moments produced by the muscle-tendon unit should equal the inverse dynamics joint moments. To guarantee the feasibility of the optimization problem, we added ideal non-physiological actuators at each joint, called reserve actuators. These actuators accounted for the joint moments that the muscle-tendon actuators could not reproduce, and their use was discouraged by penalizing their squared magnitude in the cost function. Detailed information about the formulation of the optimization problems is presented in Supplementary material B: Simulation workflows.

*2.3.4 Metabolic energy models*

Muscle metabolic rates were estimated using six metabolic energy models: Umberger et al. (UM03) (11), Bhargava et al. (BH04) (9), Houdijk et al. (HO06) (12), Lichtwark and Wilson (LW07) (13), Umberger (UM10) (14) and Uchida et al. (UC16) (35). All were based on the first law of thermodynamics, and their main features are summarized in Table 1. Briefly, muscle excitations, states, and state derivatives over a gait cycle were used to compute the total rate of metabolic energy for each muscle based on its contractile element work rate and heat rate components (Equation 2):

$$\dot{E}_i(t) = \dot{W}_{CEi}(t) + \dot{H}_{Ai}(t) + \dot{H}_{Mi}(t) + \dot{H}_{SLi}(t) \qquad (Eq.\ 2)$$

Where, for muscle $i$ at time $t$, $\dot{E}_i$ is the metabolic rate, $\dot{W}_{CEi}$ is the contractile element work rate, $\dot{H}_{Ai}$ is the heat rate due to activation, $\dot{H}_{Mi}$ is the heat rate due to maintenance, and $\dot{H}_{SLi}$ is the heat rate due to shortening and lengthening. In all models, except LW07, both heat rates depend on the muscle mass, computed based on muscle volume (see Musculoskeletal model, scaling method, and inverse kinematics and dynamics) and muscle density (11).

Eccentric muscle work could result in negative metabolic rate estimation, which would imply that muscles can absorb/gain energy during eccentric contractions. Although there is evidence that muscles can absorb heat during lengthening (36), it is physiologically questionable whether all chemical reactions are reversible when performing negative muscle work (37). In this regard, in the LW07 model, half of the contractile element work is considered to be dissipated as heat and the other half as energy absorption/gain. UM10 only includes concentric contractile element work in estimating the metabolic rate, and UC16 constrains the metabolic rate to be non-negative. The HO06 model did not describe the heat rate for eccentric contraction; nonetheless, its heat rate is typically assumed as zero (21,22), which leads to negative metabolic rates at eccentric contractions. It is also possible to estimate non-negligible negative metabolic rates in the original formulations of the UM03 and BH04 models. As negative metabolic rate is not explicitly addressed in BH04 and HO06 models, and as the UM03 model was updated to UM10 in part for this reason, we opted to further modify $\dot{H}_{SLi}$ in BH04, HO06 and UM03 when $\dot{E}_i < 0$ to assume that negative metabolic rate is dissipated by heat, as per Uchida et al. (15); $\dot{H}_{SLi}$ in Eq. 2 is replaced by $\dot{H}_{SLi,mod}$ for $\dot{E}_i < 0$ (Eq. 3):

$$\dot{H}_{SLi,mod}(t) = -\dot{W}_{CEi}(t) - \dot{H}_{AMi}(t) \qquad \text{for } \dot{E}_i(t) < 0 \qquad \text{(Eq. 3)}$$

This modification prevents the metabolic rate in the BH04, HO06, and UM03 models from becoming negative.

Primary features of the estimated energy efficiency, also called initial mechanical efficiency (38), defined as the ratio between contractile element work rate and metabolic rate, as well as the metabolic rate from different metabolic energy models, are illustrated for the soleus muscle for a general case in Fig 2. For concentric contractions, the energy efficiency in all metabolic energy models is much more influenced by the muscle length changes (velocity) than by the muscle length. The energy efficiencies of the UM03 and UM10 models are identical and highly correlated to activation level, whereas the UC16 model is less correlated with activation level except at high muscle velocities. The energy efficiencies of the BH04 and HO06 models are similar in magnitude at low to moderate rates of fiber length change and in correlation with the activation level. In contrast, the LW07 model shows the lowest energy efficiency among the models and is insensitive to activation levels. The metabolic energy rate estimated by the LW07 model was the largest, followed by the HO06, BH04, and all the models derived from Umberger et al. (11) (UM03, UM10, and UC16). Energy efficiency was not computed for eccentric contractions as it cannot be defined for most of the metabolic energy models since the negative metabolic rate is truncated. In this regard, the metabolic rate during eccentric contraction in the UM03, BH04, HO06, and UC16 models is zero, in the UM10 model is relatively low and constant across muscle velocities, and in the LW07 model yields negative values equivalent to half of the contractile element work rate (Table 1).

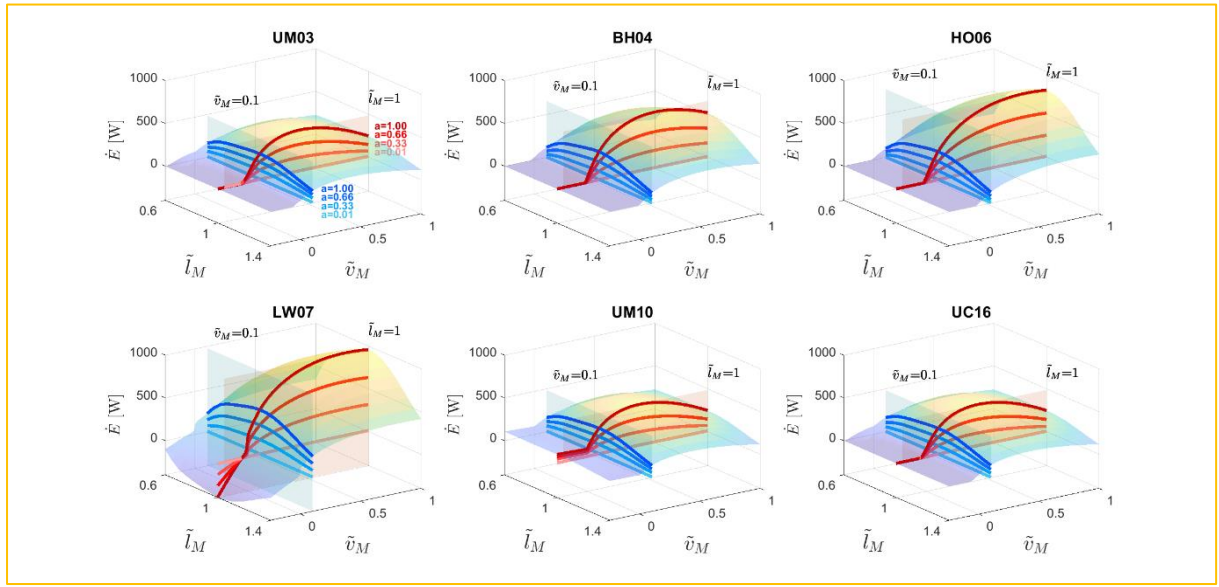

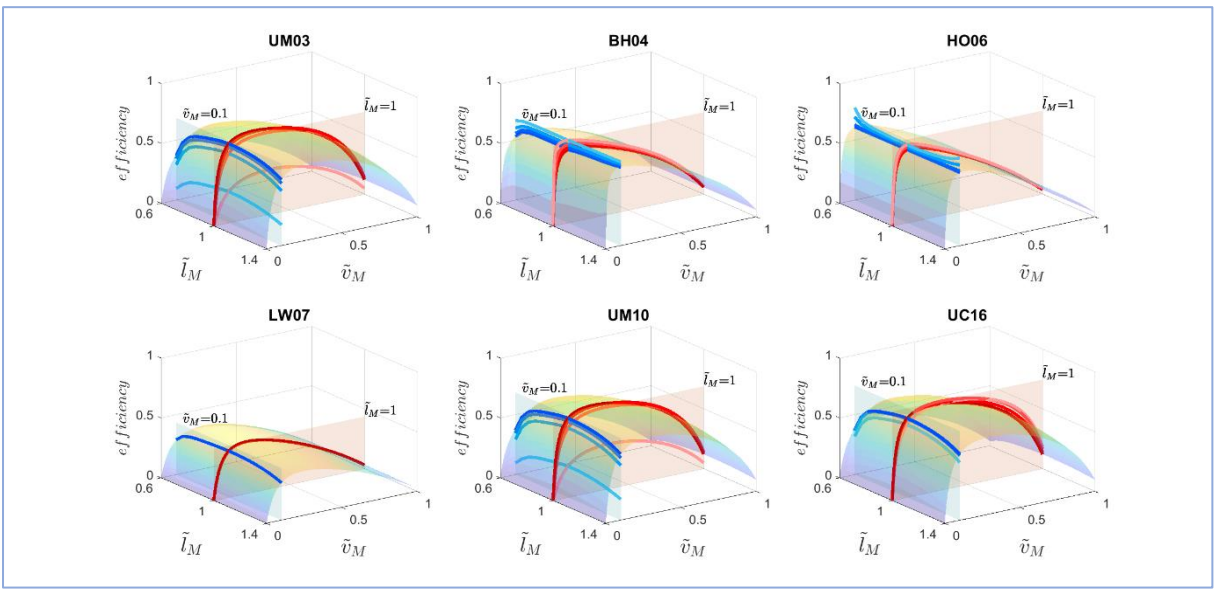

Fig 2: *Metabolic rates and energy efficiencies is soleus*. Metabolic rate ($\dot{E}$) (above) and energy efficiency (below), defined as the ratio between the contractile element work rate and metabolic rate, for a general case of the soleus at several activation levels across normalized fiber lengths ($\tilde{l}_M$), normalized fiber velocities ($\tilde{v}_M$) using six metabolic energy models: Umberger et al. (UM03), Bhargava et al. (BH04), Houdijk et al. (HO06), Lichtwark and Wilson (LW07), Umberger (UM10) and Uchida et al. (UC16). Metabolic rate and energy efficiency for activation levels 0.01, 0.33, 0.66, and 1.00 are shown for $\tilde{l}_M = 1$, (plane sections in red color gradient), and for $\tilde{v}_M = 0.1$ (plane sections in blue color gradient). Ratio of slow twitch muscle fiber, muscle mass, optimal fiber length, and maximum voluntary contraction are assumed as 0.8, 0.48 kg, 4.5 cm, and 10 [optimal fiber lengths/second], respectively.

Table 1: Principal characteristics of work rate and heat rates in six metabolic energy models: Umberger et al. (UM03), Bhargava et al. (BH04), Houdijk et al. (HO06), Lichtwark and Wilson (LW07), Umberger (UM10) and Uchida et al. (UC16). The UM03, UM10, and UC16 models had one expression that described both the activation $(\dot{H}_A)$ and maintenance $(\dot{H}_M)$ heat rates.

| Model | Work rate $(\dot{W}_{CE})$ | Activation heat rate $(\dot{H}_A)$ | Maintenance heat rate $(\dot{H}_M)$ | Shortening heat rate $(\dot{H}_S)$ | Lengthening heat rate $(\dot{H}_L)$ | Scaling parameters |
|---|---|---|---|---|---|---|
| UM03 | $F_{CE}v_M$ | $f_{UM03}(P_{S,F}, a, A_{AM})$, $\tilde{l}_M \leq 1$ <br> $f_{UM03}(f_l(\tilde{l}_M), P_{S,F}, a, A_{AM})$, $\tilde{l}_M > 1$ | | $f_{UM03}\left(P_{S,F}, a, \frac{v_M}{l_M^0}, v_M^{MAX}, A_S\right)$, $\tilde{l}_M \leq 1$ <br> $f_{UM03}\left(f_l(\tilde{l}_M), P_{S,F}, a, \frac{v_M}{l_M^0}, v_M^{MAX}, A_S\right)$, $\tilde{l}_M > 1$ | $f_{UM03}\left(P_{S,F}, a, \frac{v_M}{l_M^0}, v_M^{MAX}, A\right)$, $\tilde{l}_M \leq 1$ <br> $f_{UM03}\left(f_l(\tilde{l}_M), P_{S,F}, a, \frac{v_M}{l_M^0}, v_M^{MAX}, A\right)$, $\tilde{l}_M > 1$ | $m_M, s_E$ |
| BH04 | $F_{CE}v_M$ | $f_{BH04}(\emptyset, P_{S,F}, a_{S,F}^B)$ | $f_{BH04}(\tilde{l}_M, P_{S,F}, a_{S,F}^B)$ | $f_{BH04}(v_M, F_{CE}^{ISO}, F_M)$ | $f_{BH04}(v_M, F_M)$ | $m_M$ |
| HO06 | $F_{CE}v_M$ | $f_{HO06}(P_{S,F}, a)$ | $f_{HO06}(f_l(\tilde{l}_M), P_{S,F}, a)$ | $f_{HO06}(f_l(\tilde{l}_M), P_{S,F}, a, v_M)$ | Not described | $m_M$ |
| LW07 | $F_{CE}v_M$ | $f_{LW07}\left(f_l(\tilde{l}_M), a, v_M^{MAX}, l_M^0, F_M^0, f_V(\tilde{v}_{CE})\right)$ | | $f_{LW07}(f_l(\tilde{l}_M), a, l_M^0, F_M^0, v_M)$ | $0.5\, F_{CE} v_M$ | |
| UM10 | $F_{CE}v_M^{CON}$ | $f_{UM03}(P_{S,F}, a, A_{AM})$, $\tilde{l}_M \leq 1$ <br> $f_{UM03}(f_l(\tilde{l}_M), P_{S,F}, a, A_{AM})$, $\tilde{l}_M > 1$ | | $f_{UM03}\left(P_{S,F}, a, \frac{v_M}{l_M^0}, v_M^{MAX}, A_S\right)$, $\tilde{l}_M \leq 1$ <br> $f_{UM03}\left(f_l(\tilde{l}_M), P_{S,F}, a, \frac{v_M}{l_M^0}, v_M^{MAX}, A_S\right)$, $\tilde{l}_M > 1$ | $f_{UM03}\left(P_{S,F}, a, \frac{v_M}{l_M^0}, v_M^{MAX}, A\right)$, $\tilde{l}_M \leq 1$ <br> $f_{UM10}\left(f_l(\tilde{l}_M), P_{S,F}, a, \frac{v_M}{l_M^0}, v_M^{MAX}, A\right)$, $\tilde{l}_M > 1$ | $m_M, s_E$ |
| UC16 | $F_{CE}v_M$ | $f_{UM03}(P_{S,F}, a_{S,F}^U, A_{AM})$, $\tilde{l}_M \leq 1$ <br> $f_{UM03}(f_l(\tilde{l}_M), P_{S,F}, a_{S,F}^U, A_{AM})$, $\tilde{l}_M > 1$ | | $f_{UM03}\left(P_{S,F}, a_{S,F}^U, \frac{v_M}{l_M^0}, v_M^{MAX}, A_S\right)$, $\tilde{l}_M \leq 1$ <br> $f_{UM03}\left(f_l(\tilde{l}_M), P_{S,F}, a_{S,F}^U, \frac{v_M}{l_M^0}, v_M^{MAX}, A_S\right)$, $\tilde{l}_M > 1$ | $f_{UM03}\left(P_{S,F}, a_{S,F}^U, \frac{v_M}{l_M^0}, v_M^{MAX}, A\right)$, $\tilde{l}_M \leq 1$ <br> $f_{UM03}\left(f_l(\tilde{l}_M), P_{S,F}, a_{S,F}^U, \frac{v_M}{l_M^0}, v_M^{MAX}, A\right)$, $\tilde{l}_M > 1$ | $m_M, s_E$ |

Suffix S, F — Slow twitch $(S)$ and fast twitch $(F)$ muscle fiber, respectively

$m_M$ — Muscle mass $[kg]$

$P_{S,F}$ — Percentage of slow twitch $(P_S)$ and fast twitch $(P_F)$ fibers in the muscle [ ], respectively

$\tilde{l}_M, l_M^0$ — Normalized fiber length [ ] and optimal fiber length $[m]$, respectively

$\tilde{v}_M, v_M, v_M^{CON}, v_M^{MAX}$ — Normalized fiber velocity [ ], fiber velocity $\left[\frac{m}{s}\right]$, fiber velocity considering only concentric contraction $\left[\frac{m}{s}\right]$, and maximum contraction velocity $\left[\frac{l_M^0}{s}\right]$, respectively

$f_l(\tilde{l}_M), f_V(\tilde{v}_M)$ — Force-length and force-velocity relationships [ ], respectively.

$F_M, F_{CE}, F_{CE}^{ISO}, F_M^0$ — Muscle force, contractile element force, contractile element force considering only the force-length relationship and activation, and maximum force $[N]$, respectively

| | |
|---|---|
| $\emptyset$ | Decay function as described by Bhargava et al. 2004 [ ] |
| $A_{AM}, A_S, A$ | Scaling factor for heat production in activation and maintenance ($A_{AM}$), shortening ($A_S$), and lengthening ($A$) as described by Umberger et al. 2003 [ ] |
| $s_E$ | Scaling factor for aerobic conditions as described by Umberger et al. 2003 [ ] |
| $a$ | Muscle activation [ ] |
| $a_{S,F}^B$ | Muscle activation of slow twitch ($a_S^B$) and fast twitch ($a_F^B$) fibers based on orderly recruitment as described by Bhargava et al. 2004 [ ] |
| $a_{S,F}^U$ | Muscle activation of slow twitch ($a_S^U$) and fast twitch ($a_F^U$) fibers based on active orderly recruitment as described by Uchida et al. 2016 [ ] |

## 2.3.5. Data and statistical analysis

We evaluated the agreement between computed muscle excitations and muscle fiber lengths from each simulation workflow and available experimental observations. Specifically, we compared the computed muscle excitations of most muscles in the musculoskeletal model with digitalized values from Perry's reported fine-wire EMG observations. (39). We also compared the computed normalized fiber lengths of gastrocnemius lateralis and medialis, soleus, and vastus lateralis with reported values from ultrasound studies across walking speeds (40,41), (42,43), (44), (45). We then identified the workflow with the highest accuracy, i.e., whose computed muscle-tendon dynamics agreed overall best with experimental findings.

We evaluated metabolic energy models by comparing the experimental whole-body average metabolic rates obtained from spiroergonometry and estimated whole-body average metabolic rates from each simulation workflow. The experimental whole-body average metabolic rate was computed based on the representative (3 minutes averaged) oxygen and carbon dioxide rate values for each speed using the Brockway equation (4). The estimated whole-body average metabolic rate was computed based on the metabolic rates from metabolic energy models, plus a basal rate assumed to be 1.2 W/kg (11). The estimated whole-body average was calculated as the integral over time of the time-series metabolic rates of muscles divided by the duration of the gait cycle and multiplied by two to account for two legs (as we simulated the dynamics of one leg).

The relative costs of the stance (initial foot contact to toe-off) and swing (toe-off to next foot contact) phases were computed, normalized by the total energy cost of the gait cycle. The estimated metabolic rates of the muscle function groups at the ankle, knee, and hip joints were computed as a sum contribution of the uniarticular and biarticular muscles at each joint. For biarticular muscles, the metabolic rate was distributed between joints based on the ratio of moment arms, as implemented by Uchida et al. (15).

We used a repeated measure correlation (46) to determine whether correlations exist between estimated and experimental whole-body metabolic rates, between the estimated relative cost of the muscle groups and walking speed, and between the estimated relative cost of the stance and swing phases and walking speed. Repeated measure correlation, proposed by Bland and Altman, is suitable for establishing relationships within subjects since it adjusts for inter-individual variability (47).

# 3. Results

*3.1. Evaluation of muscle excitations, fiber lengths, and tendon stiffness estimations.*

Calibration of passive forces improved the estimation of muscle excitations of the muscle function group at the hip and knee joints compared to the generic simulation workflow, and the EMG-informed simulation workflow estimated large excitations in small muscles. Salient muscle excitation features of adductor longus, extensor digitorum longus, extensor hallucis longus, flexor digitorum longus, flexor hallucis longus, gastrocnemius medialis and lateralis, gluteus medialis, gracilis, peroneus brevis and longus, sartorius, soleus, tibialis anterior, tibialis posterior, vastus intermedius, lateralis, and medialis, were estimated with all simulation workflows (Fig 3, and S2 Fig). Excitations of the muscle function groups at the hip and knee joints – adductor magnus, bicep femoris long head, gluteus maximus, semimembranosus, and semitendinosus – were better modelled in the simulations with calibrated passive forces (EFFORT-CAL, EFFORT-TEN, and EMG-TEN), compared to simulations with generic passive forces (Fig 3, and S2 Fig). The salient feature of the rectus femoris excitation during the swing phase was better modelled with calibrated passive forces (EFFORT-CAL, EFFORT-TEN, and EMG-TEN), while the salient features of adductor brevis, iliacus, psoas excitations were not well estimated in any of the simulation workflows (S2 Fig). The EMG-informed simulation workflow (EMG-TEN) estimated a similar on/off pattern than the simulation workflow EFFORT-TEN but estimated excitation substantially higher magnitudes, especially for the rectus femoris and for small muscles around the ankle (e.g., flexor digitorum longus, S2 Fig).

Personalization of tendon stiffness improved the estimation of the normalized fiber lengths compared to the generic tendon stiffness values, and the computed fiber lengths were slightly affected by informing the muscle control with EMGs. The personalized Achilles tendon stiffness (~160 N/mm) was within the *in vivo* measured values (141 – 170 N/mm) by Stenroth et al. (48) (see Supplementary material: Simulation workflow). Also, the personalized patellar tendon estimated nearly isometric contraction of vastus lateralis during loading response with the simulation workflow EMG-TEN, which aligns with previous experimental observations (40,41). Simulation workflows without tendon stiffness personalization failed to estimate the operating range of plantarflexors as suggested in the literature, i.e., models estimated that plantarflexor muscles operated at the ascending limb or plateau (0.75 < normalized fiber length < 1.05) instead of the descending limb (normalized fiber length > 1.05) (Fig 3). Fiber lengths were not sensitive to the calibration of passive force-length curves; muscles worked at the same operating range between the simulation workflows EFFORT-GEN and EFFORT-CAL.

However, fiber lengths were affected by informing muscle excitations with EMGs. The simulation workflows EFFORT-TEN and EMG-TEN estimated somewhat different fiber lengths, though they operated in similar regions. The simulated vastus lateralis contracted concentrically and then eccentrically during the loading response in the simulation workflow EFFORT-TEN. Overall, the simulation workflow EFFORT-TEN showed the best estimation of muscle excitations and fiber lengths among all the simulation workflows; therefore, it was regarded as having the highest accuracy and was used in all subsequent analyses; results from the other simulation workflows can be found in the Supplementary Material.

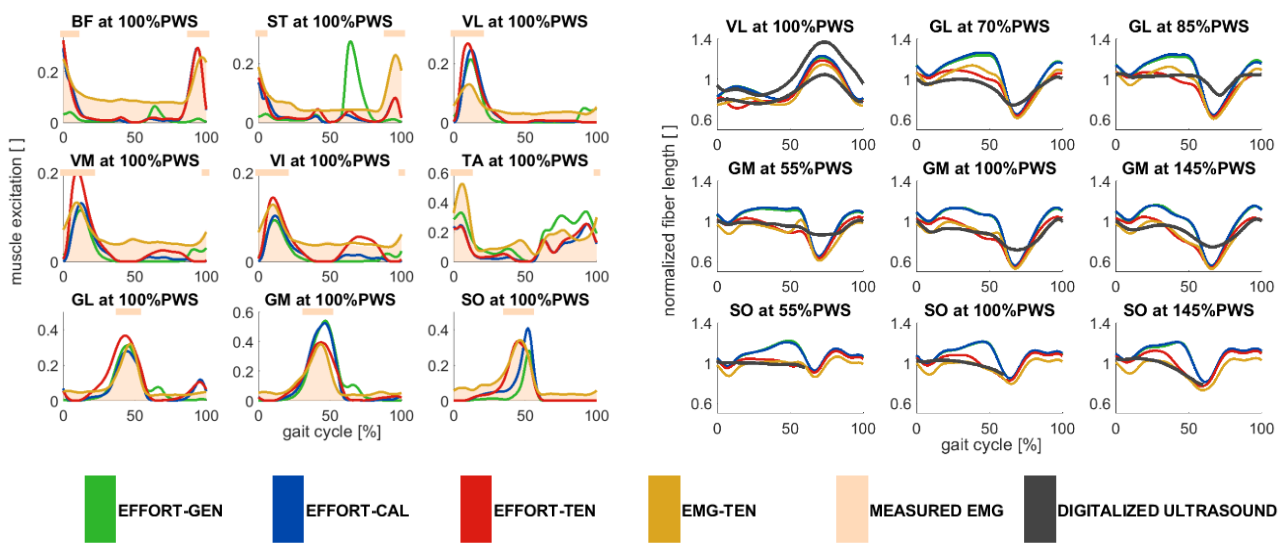

Fig 3: *Computed and experimental muscle excitations and fiber lengths*. Muscle excitations of biceps femoris long head (BF), semitendinosus (ST), vastus lateralis (VL), vastus medialis (VM), tibialis anterior (TA), gastrocnemius lateralis (GL), gastrocnemius medialis (GM) and soleus (SO) during the gait cycle at preferred walking speed (PWS) (left) and normalized fiber lengths of vastus lateralis, gastrocnemius lateralis, gastrocnemius medialis, and soleus during the gait cycle across walking speeds with four simulation workflows: Minimal muscle effort with generic passive force (EFFORT-GEN), with calibrated passive force (EFFORT-CAL), with calibrated passive force and personalized tendon stiffness (EFFORT-TEN), and EMG-informed with calibrated passive force and personalized tendon stiffness (EMG-TEN). Simulated muscle excitations and normalized fiber lengths represent the average values among all subjects. Measured EMGs represented the average values among all subjects and were scaled using optimization variables in EMG-TEN. Experimental values of fiber lengths were obtained by digitalizing previously reported experimental findings using

ultrasound in vastus lateralis (40,41), gastrocnemius lateralis (42,43), gastrocnemius medialis (44), and soleus (45). Experimental fiber lengths were normalized based on average values reported from a muscle architecture data set (49) when not otherwise reported in experimental studies. Horizontal lines above muscle excitations indicate on/off timing for EMG signals, defined as >50% excitation.

*3.2. Evaluation of whole-body average metabolic rate estimations.*

Calibration of passive forces and personalization of tendon stiffness did not substantially improve the estimation of whole-body average metabolic rate compared to the generic simulation workflow, and the EMG-informed simulation workflow overestimated metabolic rates. A linear relationship between the estimated and experimental whole-body average metabolic rate, with slope m and correlation coefficient r for all models and workflows, was analyzed for the full range of walking speeds (Fig 4). The estimated whole-body average metabolic rate correlated well with the experimental measurements with the generic simulation workflow (r≥0.93) (EFFORT-GEN); and with the inclusion of calibrated passive forces and personalized tendon stiffness (EFFORT-TEN), the correlation slightly increased (r>0.96) for all metabolic energy models. The correlation slopes were closest to 1.0 with the BH04 model (m=1.04), followed by the UM10 and HO06 models (m=0.90 and 0.86, respectively). The mean and standard deviation of the experimental whole-body average metabolic rate at the preferred walking speed was 4.5 [W/kg] and 1.1 [W/kg], respectively, which was better estimated with the UC16 model (4.5 [0.5]), followed by the HO06 and the BH04 models (4.3 [0.5] and 5.0 [0.6], respectively) (Table 2). Including EMGs in the muscle control (EMG-TEN) increased the estimated metabolic rate, led to overestimating the correlation slopes, and did not improve the correlation coefficient in most metabolic energy models. Overall, the LW07 provided the lowest agreement between estimated and experimental metabolic rates among the simulation workflows, while the BH04 model showed the highest.

Table 2: Whole-body average metabolic rate scaled by subject mass [W/kg] at preferred walking speed with four simulation workflows: Minimal muscle effort with generic passive force (EFFORT-GEN), with calibrated passive force (EFFORT-CAL), with calibrated passive force and personalized tendon stiffness (EFFORT-TEN), and EMG-informed with calibrated passive force and personalized tendon stiffness (EMG-TEN), using six metabolic

energy models: Umberger et al. (UM03), Bhargava et al. (BH04), Houdijk et al. (HO06), Lichtwark and Wilson (LW07), Umberger (UM10) and Uchida et al. (UC16). Values (mean [SD]) are computed among all subjects. The mean and standard deviation of the experimental whole-body average metabolic rate at the preferred walking speed was 4.5 [W/kg] and 1.1 [W/kg], respectively

| Models | EFFORT-GEN | EFFORT-CAL | EFFORT-TEN | EMG-TEN |
|---|---|---|---|---|
| UM03 | 6.2 [0.6] | 6.0 [0.5] | 6.4 [0.5] | 9.3 [0.7] |
| BH04 | 5.2 [0.7] | 4.8 [0.6] | 5.0 [0.6] | 7.7 [1.0] |
| HO06 | 4.5 [0.6] | 4.1 [0.5] | 4.3 [0.5] | 6.6 [0.9] |
| LW07 | 7.9 [0.9] | 7.3 [0.9] | 7.7 [0.9] | 12.2 [1.2] |
| UM10 | 7.1 [0.7] | 6.8 [0.5] | 7.2 [0.6] | 10.8 [0.9] |
| UC16 | 4.6 [0.5] | 4.3 [0.4] | 4.5 [0.5] | 6.8 [0.8] |

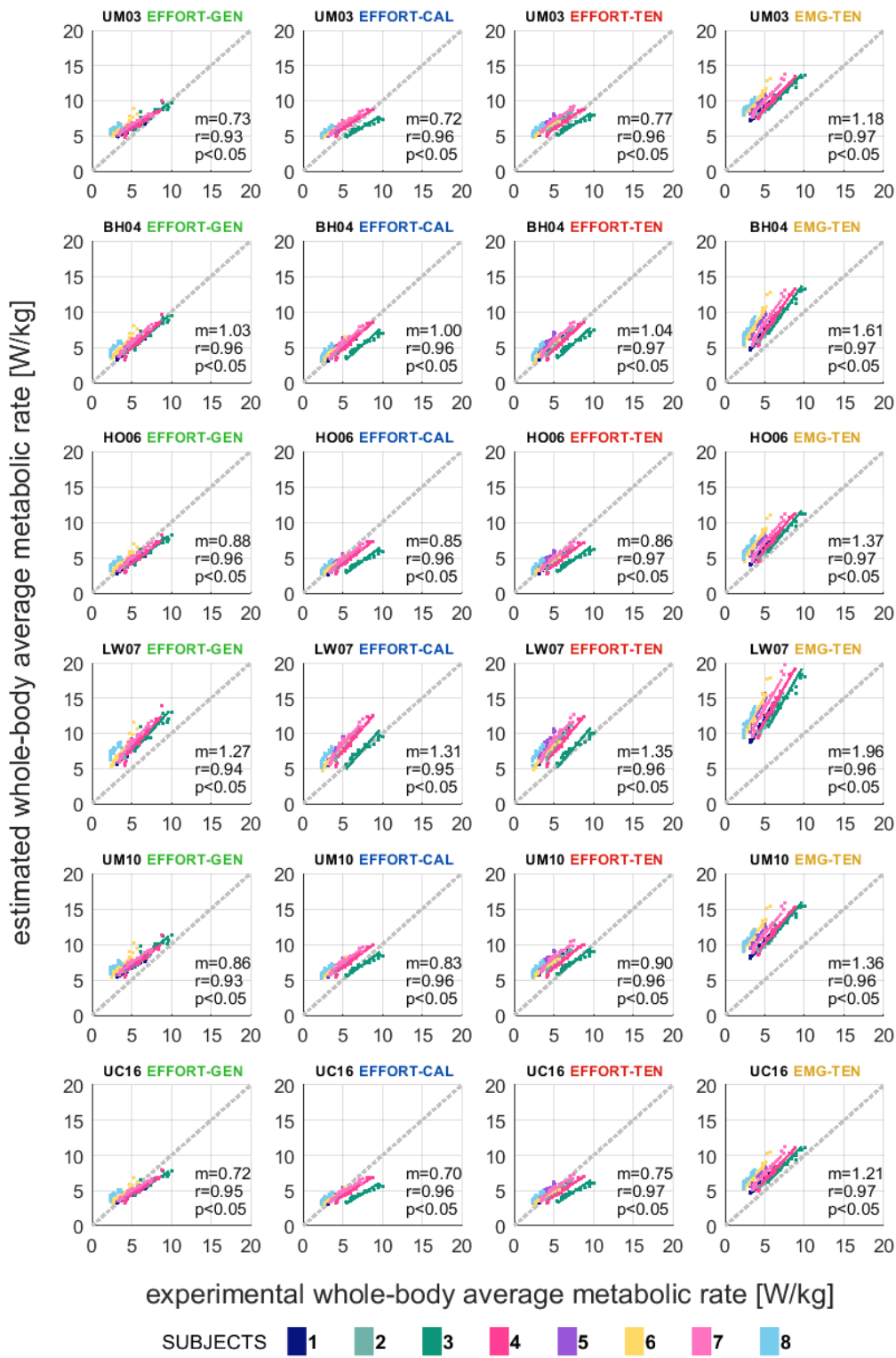

Fig 4: *Repeated measures correlation between estimated and experimental whole-body average metabolic rates*. Repeated measures correlation between estimated and experimental whole-body average metabolic rate [W/kg] over a range of metabolic demand with four simulation workflows: Minimal muscle effort with generic passive force (EFFORT-GEN), with calibrated passive force (EFFORT-CAL), with calibrated passive force

and personalized tendon stiffness (EFFORT-TEN), and EMG-informed with calibrated passive force and personalized tendon stiffness (EMG-TEN), using six metabolic energy models: Umberger et al. (UM03), Bhargava et al. (BH04), Houdijk et al. (HO06), Lichtwark and Wilson (LW07), Umberger (UM10) and Uchida et al. (UC16). Each color represents an individual subject. An ideal relationship, equivalent to $y = x$ is illustrated as a dotted line. For each simulation workflow and metabolic energy model, the slope *m*, correlation coefficient *r*, and *p*-value of the null hypothesis that no correlation between estimated and experimental metabolic rate exists are shown as means of all subjects.

*3.3. Estimation of the metabolic rates at the whole-body and muscle function group level.*

All metabolic energy models in the simulation framework with the highest accuracy estimated three metabolic rate peaks, which were more pronounced as walking increased. At the preferred walking speed, the simulation workflow with calibration of passive forces and personalization of tendon stiffness (EFFORT-TEN) estimated three metabolic rate peaks – near early stance, pre-swing and initial swing. At the muscle function group level, the first peak was mainly caused by knee and hip extensors; the second peak by plantarflexors and abductors; and the third by hip flexors, hip adductors, and knee extensors (Fig 5). In all the metabolic energy models, except with the UM10 model, the highest cost is accounted by the plantarflexors (18.4% to 24.7%), followed by the knee extensors (14.7% to 18.1%) and hip extensors (12.2% to 13%) (S1 Table). The hip adductors and abductors combined accounted for approximately 20% of the overall energy cost, and hip internal and external rotators required almost negligible cost. The levels of individualization modulated the metabolic rates, yet all metabolic energy models provided similar time-series estimates in each simulation workflow. Detailed information about the comparison between metabolic rates is presented in Supplementary material C: Comparison of metabolic estimations among simulation workflows.

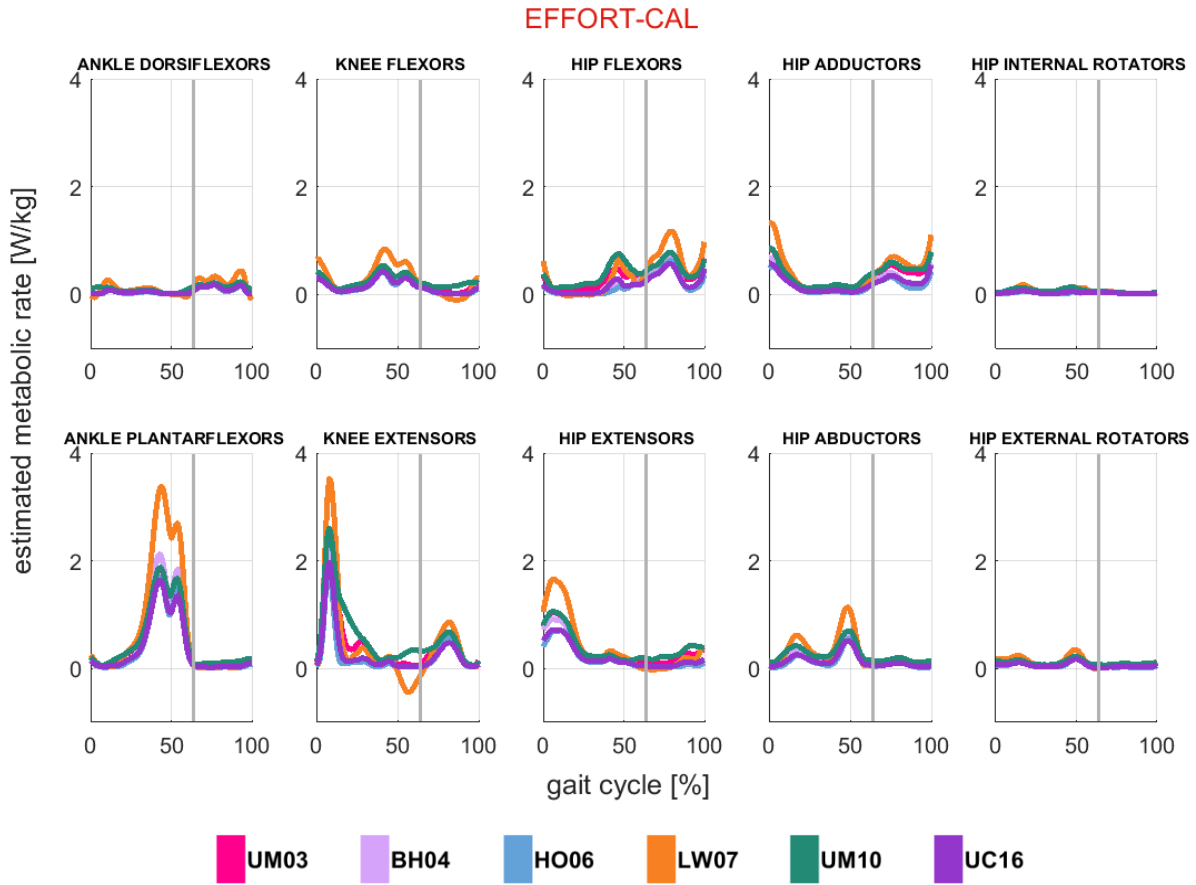

Fig 5: *Estimated metabolic rates of muscle function groups*. Estimation of metabolic rates in the muscle function groups [W/kg] at preferred walking speed with the simulation workflow based on minimal muscle effort with calibrated passive force and personalized tendon stiffness (EFFORT-TEN), using six metabolic energy models: Umberger et al. (UM03), Bhargava et al. (BH04), Houdijk et al. (HO06), Lichtwark and Wilson (LW07), Umberger (UM10) and Uchida et al. (UC16). The estimated metabolic rates represented the average values among all subjects, scaled by their mass. The vertical line represented the toe-off event.

Walking speed influenced the estimated metabolic rates throughout the gait cycle primarily by modulating the metabolic rate peaks as the speed increased. At the highest speeds, the peak near pre-swing was dominant, and at slow walking speeds, the peaks were broadly similar, wherein the peak near the early stance was lower than the others (Fig 6). Also, in all the metabolic energy models, we found a low to moderate correlation between the relative cost of the muscle function group (with respect to the total energy cost of the gait cycle) and the walking speeds (S3 Fig). In all the models, the relative cost of the ankle plantarflexors substantially increased with walking speed ($p<0.05$), while hip flexors, adductors, and abductors decreased ($p<0.05$). Trends in other muscle function groups were less pronounced (or not significant) and dependent on the metabolic energy model.

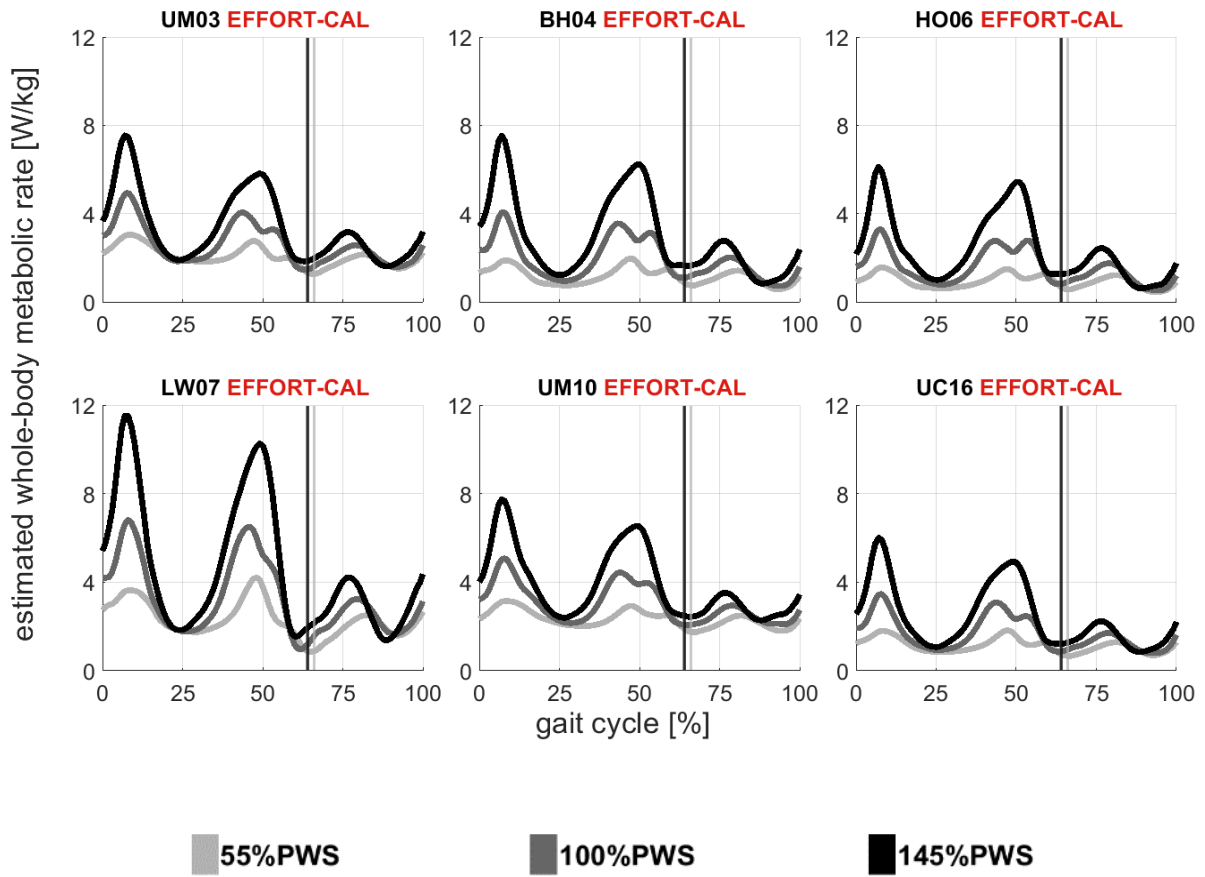

Fig 6: *Estimated whole-body metabolic rates across walking speeds*. Estimation of the whole-body metabolic rates [W/kg] at 55% PWS (light grey), 100% PWS (dark grey), and 145% PWS (black) with the simulation workflow based on minimal muscle effort with calibrated passive force and personalized tendon stiffness (EFFORT-TEN), using six metabolic energy models: Umberger et al. (UM03), Bhargava et al. (BH04), Houdijk et al. (HO06), Lichtwark and Wilson (LW07), Umberger (UM10) and Uchida et al. (UC16). The estimated metabolic rates represented the average values among all subjects, scaled by their mass. The vertical line represented the toe-off event across walking speeds.

## 3.4. Estimation of the relative costs of gait phases.

The metabolic energy models in the simulation framework with the highest accuracy estimated that the relative cost of the swing phase accounted for slightly more than one-quarter of the total energy cost of the gait cycle and also decreased with faster walking speed. At the preferred walking speed, the simulation workflow with calibration of passive forces and personalization of tendon stiffness (EFFORT-TEN) estimated a similar relative cost of the swing phase among metabolic energy models. It varied from 25.9% with the BH04 model to 29.6% with the UM10 model (Fig 7). Interestingly, these two models also provided the best estimates of whole-body average metabolic rates (Fig 4). In addition, we found a significant (p<0.05) yet moderate correlation between the relative cost of gait phases and walking speed in all the metabolic energy models (Fig 7). The correlation slope magnitudes varied from 3.57 with the UM10 model to 7.69 with the HO06 model, and was 6.87 with the BH04 model, the model that best estimated whole-body average metabolic rates. These magnitudes imply that walking speed has a low to moderate impact on the relative cost of the gait phases.

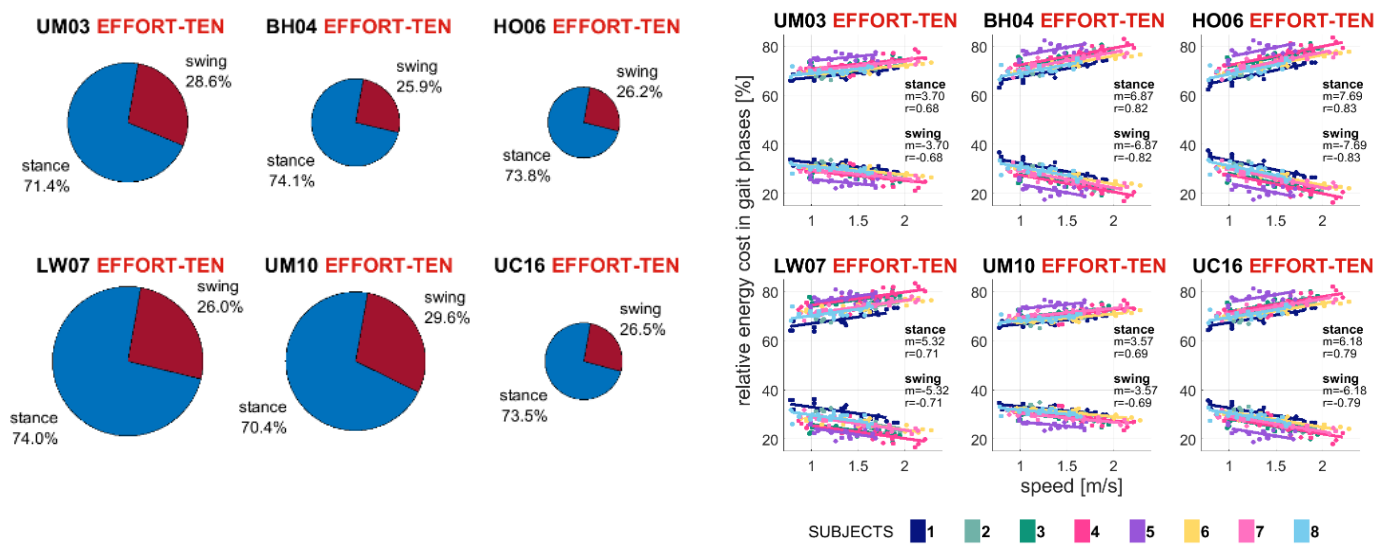

Fig 7: *Estimated relative cost of gait phases*. Cost of the stance and swing phase relative to the total energy cost in a gait cycle during preferred walking speed [left] and vs. walking speeds [right] (in percentage) with the simulation workflow based on minimal muscle effort with calibrated passive force and personalized tendon stiffness (EFFORT-TEN), using six metabolic energy models: Umberger et al. (UM03), Bhargava et al. (BH04), Houdijk et al. (HO06), Lichtwark and Wilson (LW07), Umberger (UM10) and Uchida et al. (UC16). Pie chart areas are scaled based on the total energy cost. Individual subjects are illustrated in different colors, and the slope and correlation coefficient from repeated measures correlation is indicated. The P-value was <0.05 for all correlations.

## 4. Discussion

Models with calibrated passive forces and personalized tendon stiffness better estimate muscle excitations, fiber lengths, and whole-body average metabolic rates across walking speeds compared to generic simulation workflows. Perhaps unexpectedly, models that estimate muscle excitation with information from experimentally-measured EMGs did not estimate metabolic rate accurately. This observation can be attributed to experimental limitations that make it possible to constrain only a subset of muscle EMGs. Muscle excitations are commonly validated by their agreement with EMGs (50)(51); thus, muscle excitations informed by recorded EMGs should be more reliable than muscle excitations obtained from only optimization routines. By tracking EMGs closely (+/-0.01 tracking bound), our simulation has a relatively low sensitivity to the objective function's weights. Yet, estimated excitations in most of the non-tracked muscles at the ankle and knee increased. We evaluated whether we could avoid this increment by tracking EMGs less closely (+/-0.05 tracking deviation), but found instead estimations in which the muscle-tendon states depended on the weights of the objective function, which is likely not generalizable. We simultaneously tracked recorded EMGs and minimized muscle activations, which derived into opposing goals in the objective function; see further discussion in Supplementary material B: Simulation workflows. Another alternative was to include only muscles whose recorded EMGs were available, also called EMG-driven simulations in literature (52). This approach also poses a challenge as EMGs from deep muscles cannot be recorded using surface electromyography [23]; thus, the degree to which deep muscles influence the estimations is uncertain. Furthermore, higher estimated muscle excitations led to overestimated metabolic rate in the simulation workflow EMG-TEN compared to EFFORT-TEN. This finding suggests that informing muscle control with EMGs in only a subset of muscles might propagate modeling errors that worsen estimates of muscle excitations and metabolic rate. Future work using EMG-informed simulations might benefit by adjusting temporal parameters to models of EMGs-to-force production better or by finding a better suitable trade-off between tracking EMGs and minimizing muscle effort. We determined EFFORT-TEN to have the highest accuracy in estimating muscle activations and fiber lengths among the simulation workflows and, therefore, used it for subsequent muscle metabolic rate analyses.

The BH04 model better estimated the whole-body average metabolic rate compared to the other metabolic energy models, which confirms and extends previous findings in the literature. In their comparative study of metabolic energy models to measured metabolic rate, Koelewijn et al. found higher correlation coefficients with the BH04 and the LW07 models (both r=0.96) than with the HO06 and UM03 models (r=0.94 and r=0.93, respectively), yet both models underestimated the metabolic cost, especially at high walking speeds. We found that all the metabolic energy models with the simulation workflow EFFORT-TEN estimated a high correlation (r≥0.96). We observed that the BH04 model outperformed the LW07 model in that the correlation slope was closest to 1. Arones et al. investigated the effect of musculoskeletal model personalization on estimated metabolic cost of transport in two people with post-stroke during walking using the BH04, UM03, and UM10 models (17). They found that the BH04 model with optimized muscle parameters and with muscle control computed by performing EMG-driven simulation or by minimizing muscle activation via static optimization accurately estimated trends of transportation cost. We did not perform EMG-driven simulation, yet our accurate metabolic rate estimation using the BH04 model and muscle-tendon states derived from minimal muscle effort is aligned with their findings. For the reasons mentioned above, it is suggested that the BH04 model is better suited to estimate muscle metabolic cost than other models in the literature. This suggestion might only be valid considering this study's current modeling assumptions and limitations. In fact, as Miller mentioned, the most "correct" muscle energy model is likely to predict higher metabolic rates when implemented using common assumptions in the musculoskeletal models (21). Models of the musculoskeletal system usually neglect energy attenuation due to a flexible midfoot arch and subtalar and metatarsal joint motion, for instance, which should decrease joint mechanical power (53) and, likely, metabolic demands. In this regard, accounting for such features may favor the UM10 model, which showed excellent estimation of the correlation slope yet overestimated metabolic rates. Further studies in musculoskeletal modeling might confirm or refine such observations.

Metabolic energy models estimated metabolic rates that correlated well with measured whole-body average metabolic rates, even in the generic simulation workflow, which might temp dubious conclusions about time-series metabolic rate. Our generic simulation workflow (EFFORT-GEN) using the BH04 or HO06 models estimated an almost perfect agreement with measured whole-body average metabolic rates across walking speeds, even though this simulation workflow poorly estimated excitation and fiber lengths as compared to the EFFORT-TEN simulation workflow. Furthermore, the generic simulation workflow (EFFORT-GEN) estimated a large muscle energy cost in pre-and initial swing as a product of poor modeling of passive forces, estimated a high relative muscle energy cost in the swing phase with the BH04 or HO06 models (36.5 to 36.7%, respectively), and failed to capture attenuated metabolic rate peaks due to highly compliant tendons. For further discussion, see Supplementary material C: Comparison of metabolic estimations among simulation workflows. In this regard, caution should be taken to interpret metabolic rates if the muscle activations and fiber lengths have not been validated.

Metabolic rates of the muscle function groups estimated in the simulation workflow with the highest accuracy and the BH04 and UM10 models are aligned with and extend findings from previous simulation studies. Umberger estimated that hip extensors, knee extensors, and ankle plantarflexors contributed the most to energy cost during the stance phase at a preferred walking speed (14), similar to our findings. Umberger also found that the most prominent metabolic rate peak occurred in loading response and was primarily attributed to hip extensors and hip extensors, wherein hip extensors accounted for the highest metabolic cost among muscle groups. We observed a similar timing of the metabolic rate peaks, though plantarflexors accounted for the highest metabolic cost in our simulations. In addition, we identified a metabolic rate peak during the initial swing caused by hip flexors, hip adductors, and knee extensors. In contrast, Umberger observed little-to-no contribution to metabolic rate from the knee extensors, particularly the rectus femoris. Discrepancies in metabolic rates are likely to be attributed to a higher muscle activation of the hip and knee extensors at heel strike compared to our simulation and the lack of rectus femoris activation during the initial swing estimated by Umberger. We estimated a rectus femoris activation peak at the initial swing, which is consistent with experimental observation (29,54). Interestingly, Umberger estimated that the relative cost of the swing phase is 29% which is nearly identical to our estimations with the UM10 model. Mohammadzadeh et al., using the UM10 model with modifications of Uchida et al. (15), estimated that plantarflexors accounted for 57% of the total metabolic cost, the highest among muscle function groups, followed by knee and hip extensors with 22% and 11%, respectively, and the ankle dorsiflexors and knee flexors, combined, with only 10% during walking at 1 m/s (55). While we also identified that, among muscle function groups, the plantarflexors demanded the highest metabolic cost, even at low speeds, our estimated costs are approximately 1/3 of the reported magnitudes. The muscle actuators' low fidelity to reproducing recorded knee and hip flexor joint moments might explain the discrepancy. In these reported studies, walking was simulated with muscle-tendon actuators driven by recorded EMGs and optimized fiber and tendon lengths, wherein knee and hip flexors generated little to no moment compared to the inverse dynamics solution, which resulted in overestimated metabolic cost of ankle plantarflexors and knee extensors and failure to capture the metabolic cost of the hip flexors, the primary muscle function group for the swing phase, according to our estimations. Similarly, Markowitz and Herr simulated walking using muscle-tendon actuators driven by recorded and digitalized EMGs and optimized muscle-tendon parameters to estimate muscle forces and metabolic rate using the UM10 model (25). They estimated that the swing phase accounted for 26% of the total cost of walking, identical to our estimations with the BH04 model and slightly lower than the UM10 model. In addition, we estimated a substantial cost of the hip adductor and abductors, nearly 20%, which was not described in the previous studies as they simulated hip articulation in the sagittal plane only. These comparisons suggest that estimated metabolic rates in muscle function groups might be obscured by inadequate accuracy of modeled muscle-tendon mechanics and that the swing phase accounts for 26 to 29% of the total energy cost of the gait cycle. Also, we showed that the relative cost of gait phases is sensitive to walking speed. We estimate that the relative cost of the swing phase varies between 3.6 to 6.9% per meter/second during walking using the UM10 and BH04 model, respectively. Future experimental studies might bring more insights to refine the estimates of metabolic cost.

In summary, evaluating the metabolic rates of muscles supported by various degrees of experimental observation provides insights into the relationship between muscle-tendon function and metabolic demands and suggests venues to improve our understanding of metabolic rates. We recommend adequately modeling passive structures, specifically highly compliant tendons and passive force curves, to describe better time-series trends of muscle activations, fiber lengths, and, thus, metabolic rates. Metabolic energy models that account for heat rates based on fiber-type composition and null or small eccentric contraction cost, such as in the BH04 and UM10 models, better describe whole-body metabolic rates. Finally, our observations lead to the following suggestions:

- The most effective strategy to improve metabolic efficiency is to support ankle plantarflexors during pre-swing, knee and hip extensors during loading response, and hip flexors during swing; strength training or assistive devices that target these functions have the greatest potential to reduce metabolic cost and increase efficiency.

- Among muscle function groups, the ankle plantarflexors account for the highest metabolic cost across walking speeds. Also, plantarflexor activity becomes substantially more costly when walking speed increases, whereas knee and hip extensor activity is proportionally unaffected by walking speed. Ankle plantarflexors should therefore be the primary target for support to achieve efficient walking at high speeds.

- For metabolic efficiency, it is relatively more important to assist muscle function groups that support the swing phase, specifically hip flexors and adductors, at low rather than fast walking speeds. Assistive devices that support hip flexors have been reported to significantly reduce the metabolic rates near the preferred walking speed (56). We anticipate that assistive devices that support hip flexors have the greatest potential to reduce metabolic cost at slow walking speeds.

- Hip abductors and adductors contribute to approximately one-fifth of the metabolic cost at preferred walking speed, thus devices that support and assist frontal plane hip muscles seemed recommended for walking efficiency. Moreover, supporting these muscles is relatively more important at low rather than high walking speeds.

- Rehabilitation treatments aiming to improve walking efficiency after trauma or clinical interventions should be targeted to support hip abductors, adductors, and flexion functions at slow walking speed, and then progressively shifted to support ankle plantarflexors as walking speed increases.

## 5. Data availability

Experimental data to replicate this study, such as subject anthropometrics, marker trajectories, ground reaction forces, electromyographic signals, and metabolic rates, as well as the results of the simulation workflows, will be available in a public repository.

## 6. Code availability

The scripts for calibrating the passive forces, performing the simulation workflows, and computing the metabolic rates based on the metabolic energy models are available in the following repository: https://github.com/israelluis/MetabolicRateWalking

## 7. Author contributions

I.L. and E.M.G.F conceived the study concept and design. I.L. performed the data collection, processed the data, and drafted the manuscript. I.L. and M.A. set up the simulation workflows. I.L., M.A., F.D., and E.M.G.F. analyzed and interpreted the results and edited the manuscript. E.M.G.F. obtained funding and supervised the study.

## 8. Competing interests

The authors declare no competing interests.

## 9. Materials & Correspondence

Correspondence and requests for materials should be addressed to I.L.

## 10. Acknowledgment

Authors acknowledge the funding sources provided by the Swedish Research Council (nr 2018-00750) and Promobilia Foundation (nr 18200).